\documentclass[aps,prx,onecolumn,unsortedaddress,nofootinbib,floatfix]{revtex4-2}
\usepackage{amsthm}
\usepackage{amsfonts}
\usepackage{siunitx}
\usepackage{amsmath}
\usepackage{amssymb}
\usepackage{graphicx}
\usepackage{verbatim}
\usepackage[colorlinks]{hyperref}
\usepackage{tikz}
\usepackage{braket}
\usepackage{xcolor}
\usepackage{setspace}
\newcommand{\aqpu}{A_\text{QPU}}
\usepackage{setspace}

\makeatletter
\def\l@subsection#1#2{}
\def\l@subsubsection#1#2{}
\makeatother

\definecolor{linkcolor}{RGB}{0,83,166}
\hypersetup{
  colorlinks = true,
  allcolors = {linkcolor}
}

\makeatletter
  \@addtoreset{section}{part}
  \makeatother 

\begin{document}
\newcommand{\mytitle}{Tutorial: Calibration refinement in quantum annealing}
\title{\mytitle}

\newcommand{\affildw}{D-Wave, Burnaby, British Columbia, Canada}
\newcommand{\affilsfu}{Department of Physics, Simon Fraser University, Burnaby, British Columbia, Canada}
\author{Kevin Chern, Kelly Boothby, Jack Raymond,  Pau Farr\'{e}, and Andrew D.~King}

\affiliation{\affildw}
\date{\today}

\begin{abstract}
  Quantum annealing has emerged as a powerful platform for simulating and optimizing classical and quantum Ising models.  Quantum annealers, like other quantum and/or analog computing devices, are susceptible to nonidealities including crosstalk, device variation, and environmental noise.  Compensating for these effects through calibration refinement or ``shimming'' can significantly improve performance, but often relies on ad-hoc methods that exploit symmetries in both the problem being solved and the quantum annealer itself.  In this tutorial we attempt to demystify these methods.  We introduce methods for finding exploitable symmetries in Ising models, and discuss how to use these symmetries to suppress unwanted bias.  We work through several examples of increasing complexity, and provide complete Python code.  We include automated methods for two important tasks: finding copies of small subgraphs in the qubit connectivity graph, and automatically finding symmetries of an Ising model via generalized graph automorphism.  \makebox{Code is available at \url{https://github.com/dwavesystems/shimming-tutorial}.}
\end{abstract}
\maketitle

\def\title#1{\gdef\@title{#1}\gdef\THETITLE{#1}}

\tableofcontents

\newpage
\part{Background}

\section{Introduction to quantum annealing}

Quantum annealing (QA) \cite{Kadowaki1998, Johnson2011} is a computing approach that physically realizes a system of Ising spins in a transverse magnetic field.  A common application of QA is to find low-energy spin states of the Ising problem Hamiltonian
\begin{equation}\label{eq:hamc}
  \mathcal H_P = \sum_{i} h_i\sigma^z_i + \sum_{i<j} J_{ij}\sigma^z_i\sigma^z_j.
\end{equation}
Here, $\{\sigma^z_i\}_{i=1}^N \in \{-1,1\}^N$ is a set of Pauli $z$-operators, which can be thought of as a vector of classical $\pm 1$ Ising spins; $h_i$ denotes a longitudinal field (bias) on spin $i$, and $J_{ij}$ (used interchangeably with $J_{i,j}$ depending on context) denotes a coupling (quadratic interaction) between spins $i$ and $j$.  Minimizing $\mathcal H_P$ is intractable, i.e., NP-hard \cite{Barahona1982}.

\begin{figure}[b]
  \includegraphics[scale=0.8]{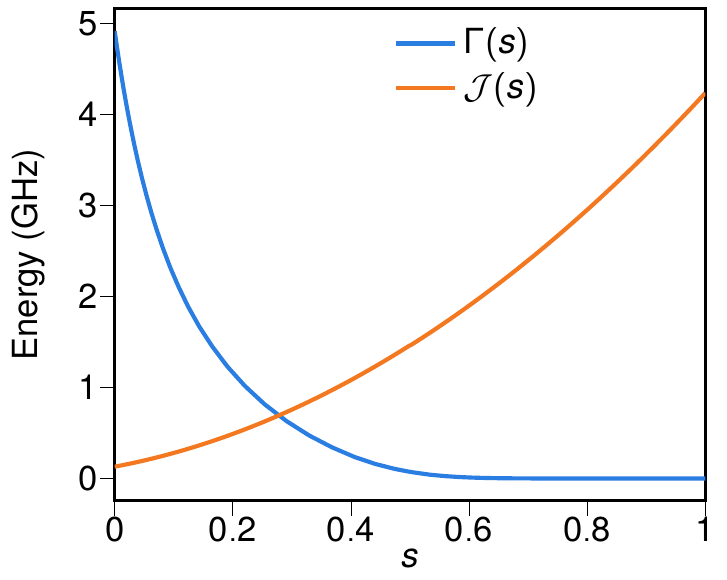}
  \caption{Annealing schedule for Hamiltonian \ref{eq:ham} in a D-Wave\texttrademark\ Advantage\texttrademark\ processor.  $\Gamma(s)$ and $\mathcal J(s)$ control the magnitude of quantum fluctuations and the Ising energy scale, respectively.  These values vary slightly from one processor to another.}\label{fig:sched}
\end{figure}

QA adds to $\mathcal H_P$ an initial driving Hamiltonian
\begin{equation}\label{eq:hami}
  \mathcal H_D = -\sum_{i}\sigma^z_i.
\end{equation}
The ground state of $\mathcal H_D$, which is a uniform quantum superposition of all classical states, is easy to prepare.  QA guides a time-dependent Hamiltonian $\mathcal H(s)$ from $\mathcal H_D$ to $\mathcal H_P$, by linearly combining $\mathcal H_D$ and $\mathcal H_P$ as
\begin{equation}\label{eq:ham}
  \mathcal H(s)= \Gamma(s) \mathcal H_D + \mathcal J(s) \mathcal H_P,
\end{equation}
where $s$ is a unitless annealing parameter ranging from $0$ to $1$.  Unless stated, $s$ is simply $t/t_a$: time normalized by annealing time.  The functions $\Gamma(s)$ and $\mathcal J(s)$ define the {\em annealing schedule}: $\Gamma(s)$ decreases toward $0$ as a function of $s$, and $\mathcal J(s)$ increases as a function of $s$; $\Gamma(0)\gg \mathcal J(0)$.  An example is shown in Fig.~\ref{fig:sched}.

\section{Calibration imperfections and refinement}

Quantum processing units (QPUs, in this case quantum annealers) are typically made available with a single one-size-fits-all calibration.  Nonidealities in the calibration can arise from a number of sources.  For example, small fluctuations in the magnetic environment can bias qubits in one direction or the other.  So can {\em crosstalk}, in which a Hamiltonian term, e.g., a programmed coupler $J_{ij}$, can cause an undesired perturbation in another Hamiltonian term corresponding to a physically nearby device, e.g.\ a bias field $h_i$.

In short, no calibration is perfect.  Oftentimes, in-depth studies of a single system (Ising model) or ensemble of systems (e.g., a set of realizations of a spin-glass model) can be improved by suppressing crosstalk and other nonidealities.  This is achieved by ``shimming'': inferring statistical features of an ideal annealer, and tuning the Hamiltonian to produce these features.  An ideal annealer, in this work, is defined simply as one that respects symmetries in the Hamiltonian---each qubit behaves identically, and each coupler behaves identically.

Variations on the methods described herein have by now been used in many works \cite{King2018,Kairys2020,Nishimura2020,King2021a,King2021,QubitSpinIce,King2022,King2022b}.  Often, when behavior of the system relies on precise maintenance of energy degeneracy between states, or energy splitting from the transverse field, results are highly sensitive to these tunings.  Particularly for the simulation of exotic magnetic phases, calibration refinement is an essential ingredient of successful experiments.  Yet, so far the discussion of these methods has mostly been relegated to supplementary materials.  Our aim here is to provide an accessible guide that will encourage the use of these powerful but simple methods.

Specific visual demonstrations of the benefit of these methods ``in the wild'' include:
\begin{itemize}
  \item Frustrated 2D lattice, \cite{King2018} Extended Data Fig.~7.
  \item Diluted ferromagnet, \cite{Nishimura2020}, Figs.~34--35.
  \item 1D quantum Ising chain, \cite{King2022}, Supplementary Materials, Figs.~S3--S4.
  \item 3D quantum spin glasses, \cite{King2022b}, Supplementary Materials, Figs.~S8--S9.
\end{itemize}

\section{Inferring statistical features: Qubit and coupler orbits}\label{sec:orbits}

\begin{figure}
  \includegraphics[scale=0.6]{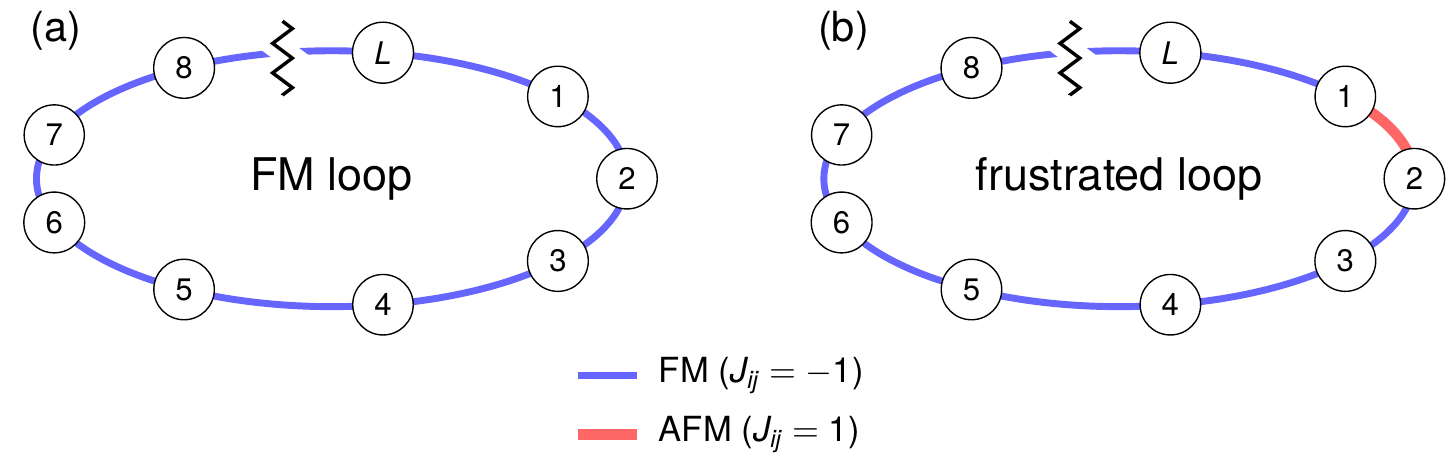}
  \caption{(a) Ferromagnetic loop (periodic 1D chain) on $L$ spins. (b) Frustrated loop, with one antiferromagnetic coupler.  The FM loop has a twofold-degenerate ground state (all spins up or all spins down) with no frustration; the frustrated loop has $2L$ ground states, each with one frustrated bond.  When $h=0$, all qubits trivially have zero average magnetization in an ideal annealer.}\label{fig:chains}
\end{figure}

The approach described in this tutorial can be stated simply and generically: {\bf In theory, two observables of a QPU output are expected to be identical due to symmetries in the Ising model being studied.  In experiment they can differ systematically.  We tune Hamiltonian terms to reduce these differences.}

In this work we only consider one- and two-spin observables---spin magnetizations and frustration probabilities---in part because they can be fine-tuned easily using the available programmable terms in the QPU.  A call to the QPU typically results in a number of classical samples, which we set to 100 for all examples.  From these samples we can compute a magnetization
\begin{equation}m_i = \langle s_i\rangle \in [-1,1]
\end{equation}
for each spin $s_i$, and a {\em frustration probability}
\begin{equation}
  f_{i,j}= \frac{1 + \text{sign}(J_{i,j})\langle s_is_j\rangle }{2} \in [0,1]
\end{equation}
for each coupler $J_{i,j}$; $f_{i,j}$ is the observed probability of the coupler having a positive contribution to the energy in $\mathcal H_P$.

This raises the first question: how do we identify observables that should be identical in expectation?  The answer is: through symmetries of the Ising model under spin relabeling and gauge transformation\footnote{A gauge transformation is also known as a {\it spin reversal transformation}, in which a subset of spins have their sign flipped.}.  We understand and formalize these symmetries---and automate their detection---through {\em graph isomorphisms} (especially automorphisms) and generalizations thereof~\cite{Godsil2001}.  The symmetries we find and exploit here are a subset of all possible symmetries.

These symmetries admit two types of equivalence relations on an Ising model $\mathcal H_P$: one on the qubits, and one on the couplers.  We call the equivalence classes {\em qubit orbits} and {\em coupler orbits}, respectively.  We use notation $\mathcal O(s_i)$ for a qubit orbit containing spin $s_i$, and $\mathcal O(s_i,s_j)$ for a coupler orbit containing coupler $(s_i,s_j)$.  We define them as having the following properties guaranteed by symmetry in an ideal annealer:
\begin{itemize}
  \item All qubits in the same orbit have the same expected magnetization.
  \item All couplers in the same orbit have the same expected frustration probabilities.
\end{itemize}
Due to spin-flip symmetries, each qubit and coupler orbit can additionally have up to one nonempty orbit that is {\it opposite}.
\begin{itemize}
  \item If qubit orbits $\mathcal O(s_i)$ and $\mathcal O(s_j)$ are opposite, then
        \begin{itemize}
          \item We write $\mathcal O(s_i) = -\mathcal O(s_j)$ and $-\mathcal O(s_i) = \mathcal O(s_j)$.
          \item If $\mathcal O(s_i) = -\mathcal O(s_j)$, then $h_i=-h_j$ and, in an ideal annealer, $m_i=-m_j$.
        \end{itemize}
  \item If coupler orbits $\mathcal O(s_i,s_j)$ and $\mathcal O(s_k,s_\ell)$ are opposite, then
        \begin{itemize}
          \item We write $\mathcal O(s_i,s_j) = -\mathcal O(s_k,s_\ell)$ and $-\mathcal O(s_i,s_j) = \mathcal O(s_k,s_\ell)$.
          \item $J_{i,j} = J_{k,\ell}$ and, in an ideal annealer, $f_{i,j} = f_{k,\ell}$.
        \end{itemize}
\end{itemize}
We will sometimes overload notation, conflating $\mathcal O(s_i)$ with $\mathcal O(i)$, and $\mathcal O(s_i,s_j)$ with $\mathcal O(i,j)$.

Qubit and coupler orbits are related to, but not identical to, automorphism orbits of an auxiliary graph.  In particular, qubit and coupler orbits are not unique: Putting each qubit and each coupler in a separate orbit is sufficient to meet the definition, but does not provide any useful information.  We seek large orbits that satisfy the requirements.

Note that in the commonly arising situation where $h_i=0$ on all qubits, each qubit orbit is its own opposite, so all qubits have $m_i=0$.  The analogous situation does not exist for couplers, because we do not consider symmetries between pairs of qubits with zero coupling between them.  Two simple examples are shown in Fig.~\ref{fig:chains}: a frustrated loop and an unfrustrated loop.  In each case, all qubits are expected to have magnetization and all couplers are expected to have the same probability of frustration, but this is less obvious in the frustrated case than in the ferromagnetic case.

Having defined qubit and coupler orbits, we now consider how to find them.

\subsection{Automorphisms of the signed Ising model}
\begin{figure}[t]
  \includegraphics[scale=1]{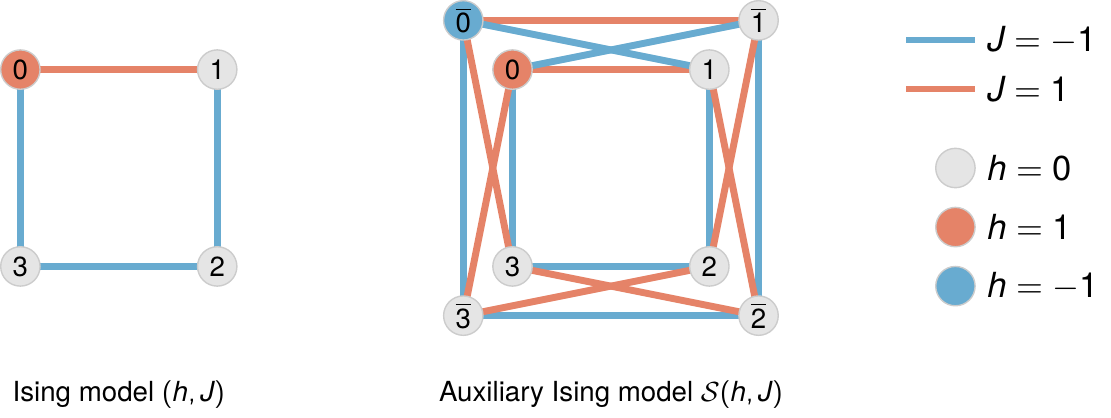}
  \caption{{\bf Construction of signed Ising model}. To detect exploitable symmetries, we search for automorphisms of an auxiliary Ising model in which each spin is duplicated into itself and its negation; each coupler is then expanded to four copies of itself, two of them negated.  Automorphisms of the auxiliary Ising model can be detected by conversion into an equivalent automorphism-finding problem on an edge-labeled graph.  Here, vertex labels indicate the identities of spins, and show how each spin is duplicated for the signed Ising model.}\label{fig:auxiliary1}
  \vspace{2em}\includegraphics[scale=1]{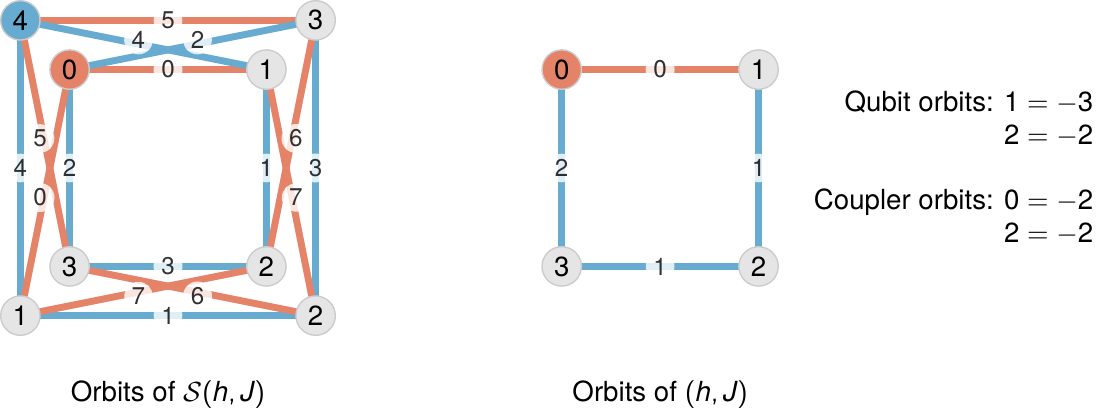}
  \caption{{\bf Orbits of signed and original Ising model}.  By computing automorphism groups of the edge- and vertex-labeled graph of the signed Ising model (Fig.~\ref{fig:auxiliary1}, right) we can construct orbits of qubits and couplers that should behave identically by symmetry in $\mathcal S(h,J)$ (left).  Here, vertex and edge labels indicate orbits.  By identifying equivalent orbits (e.g., coupler orbits 0 and 5) and reducing back to the original Ising model $(h,J)$, we determine effective qubit and coupler orbits of $(h,J)$, and their opposite relations (right).}\label{fig:auxiliary2}
\end{figure}

Let $(h,J)$ denote an Ising model with fields $h=\{h_i|v_i\in V\}$ and $J=\{J_{i,j} | e_{i,j}\in E\}$, with an underlying graph $G=(V,E)$ with vertex and edge sets $V$ and $E$.  We construct a {\it signed Ising model} $\mathcal S(h,J)$ as follows:
\begin{itemize}
  \item For each spin $v_i\in V$, $\mathcal S(h,J)$ has two spins $v_i$ and $\bar v_i$, with fields $h_i$ and $-h_i$ respectively.
  \item For each coupler $e_{i,j}=(v_i,v_j) \in E$, $\mathcal S(h,J)$ has four couplers: two couplers $(v_i,v_j)$ and $(\bar v_i,\bar v_j)$ with coupling $J_{i,j}$, and two couplers $(\bar v_i,v_j)$ and $(v_i,\bar v_j)$ with coupling $-J_{i,j}$.
\end{itemize}
Informally, we simply replace each spin with two: itself and its negation, and replace each coupler with four couplers with appropriate parity-based sign flipping.  Fig.~\ref{fig:auxiliary1} shows an example of this construction applied to a four-spin Ising model.

Our aim is to find large qubit and coupler orbits for $(h,J)$, and we will begin by finding the automorphism group of $\mathcal S(h,J)$, which can be considered as a vertex- and edge-labeled graph.  The automorphism group of $\mathcal S(h,J)$ naturally generates one equivalence relation defining qubit orbits, and another equivalence relation defining coupler orbits (see Fig.~\ref{fig:auxiliary2})\footnote{Since the automorphism-finding code {\tt nauty} \cite{McKay2014} only handles vertex-labeled graphs and not edge-labeled graphs, we need to construct a vertex-labeled graph $G''$ from $\mathcal S(h,J)$ which gives us the appropriate automorphism group.}.  Our orbits of $\mathcal S(h,J)$ immediately give us orbits of $(h,J)$, constructed by simply discarding the qubits and couplers that do not exist in $(h,J)$.

There is more usable information held in the orbits of $\mathcal S(h,J)$.  First, we can combine coupler orbits of $\mathcal S(h,J)$ such that for each coupler $e_{i,j}\in E$, $(\bar v_i,v_j)$ and $(v_i,\bar v_j)$ are in the same orbit, and  $(v_i,v_j)$ and $(\bar v_i,\bar v_j)$ are in the same orbit.  Second, we can then easily derive opposite orbits: $\mathcal O(v_i) = -\mathcal O(\bar v_i)$, and $\mathcal O(v_i,v_j) = -\mathcal O(\bar v_i,\bar v_j)$.

These orbits are already very useful, but we can combine some to make even larger orbits.  As demonstrated in the example in Fig.~\ref{fig:auxiliary1}, in $\mathcal S(h,J)$ the couplers between pairs $(\bar v_i,v_j)$ and $(v_i,\bar v_j)$ are not necessarily automorphic.  Yet they are clearly equivalent under a flip of all spins.   Thus we combine the coupler orbits containing these two couplers.  Likewise, the same applies to couplers between pairs  $(\bar v_i,\bar v_j)$ and $(v_i, v_j)$.  This is all demonstrated in the accompanying code \href{https://github.com/dwavesystems/shimming-tutorial/tree/main/tutorial_code}{\tt example0\_1\_orbits.py}, and shown in Fig.~\ref{fig:auxiliary2}.

We now consider how to exploit orbits to improve performance in quantum annealers, building up a set of tools in the following worked examples.

\clearpage
\part{Worked Example: Ferromagnetic loop}

\begin{figure}[h]{\includegraphics[scale=0.8]{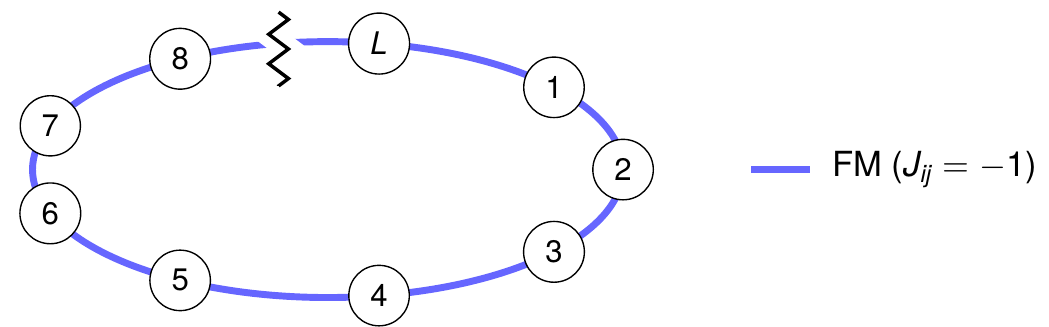}}
  \caption{Ferromagnetic loop.}
\end{figure}

\noindent{\bf Code reference:} \href{https://github.com/dwavesystems/shimming-tutorial/tree/main/tutorial_code}{\tt example1*.py}.

For our first example of calibration refinement, we study the ferromagnetic loop in which each coupling $J_{i}=J_{i,i+1}$ is equal and each field $h_i$ is zero.  In this case, by rotation, it is obvious that all qubits are in the same orbit and all couplers are in the same orbit.  Furthermore, the orbit containing all qubits is its own opposite.  Thus we will perform two refinements:  First, we will balance each qubit at zero magnetization $m_i \approx 0$.  Second, we will balance the couplings so that each coupler is frustrated with approximately equal probability.

Since the FM loop has no frustrated bonds in the ground state, the latter condition is only interesting if we sample excited states.  To ensure abundant excitations, we study a reasonably long loop with weak couplings: $L=64$ and $J_{ij}=-0.2$.

\section{Finding multiple embeddings of a small Ising model}

\noindent{\bf Code reference:} \href{https://github.com/dwavesystems/shimming-tutorial/tree/main/tutorial_code}{\tt embed\_loops.py}.

The first task is to find a copy of the FM loop in the qubit connectivity graph $\aqpu$ of the QPU being used.  This is an {\it embedding}---a mapping of spins of an Ising model to qubits in a QPU. In an Advantage processor, a 64-qubit loop can be embedded many times on disjoint sets of qubits, so we can run many copies in parallel for a richer and larger set of measurements.

To find these embeddings we use the Glasgow graph solver \cite{McCreesh2020}, which has been incorporated into the embedding finding module {\tt minorminer} \cite{minorminer}.  To make the embedding search faster, we raster-scan across $2\times 2$ blocks of unit cells in the QPU's Pegasus graph \cite{Boothby2020}, then greedily construct a large set of non-intersecting embeddings.  The file \href{https://github.com/dwavesystems/shimming-tutorial/tree/main/tutorial_code}{\tt embed\_loops.py} provides a code example that finds multiple disjoint copies of a 64-qubit loop in $\aqpu$.

\section{Balancing qubits at zero}
\noindent{\bf Code reference:} \href{https://github.com/dwavesystems/shimming-tutorial/tree/main/tutorial_code}{\tt example1\_1\_fm\_loop\_balancing.py}.

We will use simple parameters for the experiment, running \SI{1}{\micro s} anneals and drawing 100 samples for each QPU call.  We set {\tt auto\_scale=False} to ensure that the QPU will not automatically magnify the energy scale.

In D-Wave's annealing QPUs, each qubit $s_i$ can be biased toward $-1$ or $+1$ in two ways: first, with a programmable longitudinal field $h_i$ as in Eq.~\ref{eq:hamc}; second, with a programmable flux-bias offset (FBO) $\Phi_i$ \cite{fbo, Harris2009}.  In the quantum annealing Hamiltonian (\ref{eq:ham}), the bias conferred by the $h_i$ term is scaled by $\mathcal J(s)$, meaning that it changes as a function of $s$.  The FBO $\Phi_i$, in contrast, confers a constant bias that is independent of $s$.  We prefer to mitigate biases using FBOs, in part because they are programmed independently of $h_i$.

We employ an iterative gradient descent method for minimizing $|m_i|$ with a step size $\alpha_\Phi$.  For a given iteration we consider the observed magnetization $m_i = \langle s_i\rangle$.  If $m_i<0$ we adjust the FBO to push $s_i$ toward $+1$; if $m_i>0$ we adjust the FBO to push $s_i$ toward $-1$.  This is done by updating
\begin{equation}\label{eq:fboshim}
  \Phi_i \leftarrow \Phi_i -\alpha_\Phi(m_i-\bar m)
\end{equation}
for each qubit after each iteration, where $\bar m$ is the average observed magnetization across all qubits.  In this case, we can simply replace $\bar m$ with $0$ since $h_i=0$ for all qubits.

\begin{figure}
  \includegraphics[scale=1]{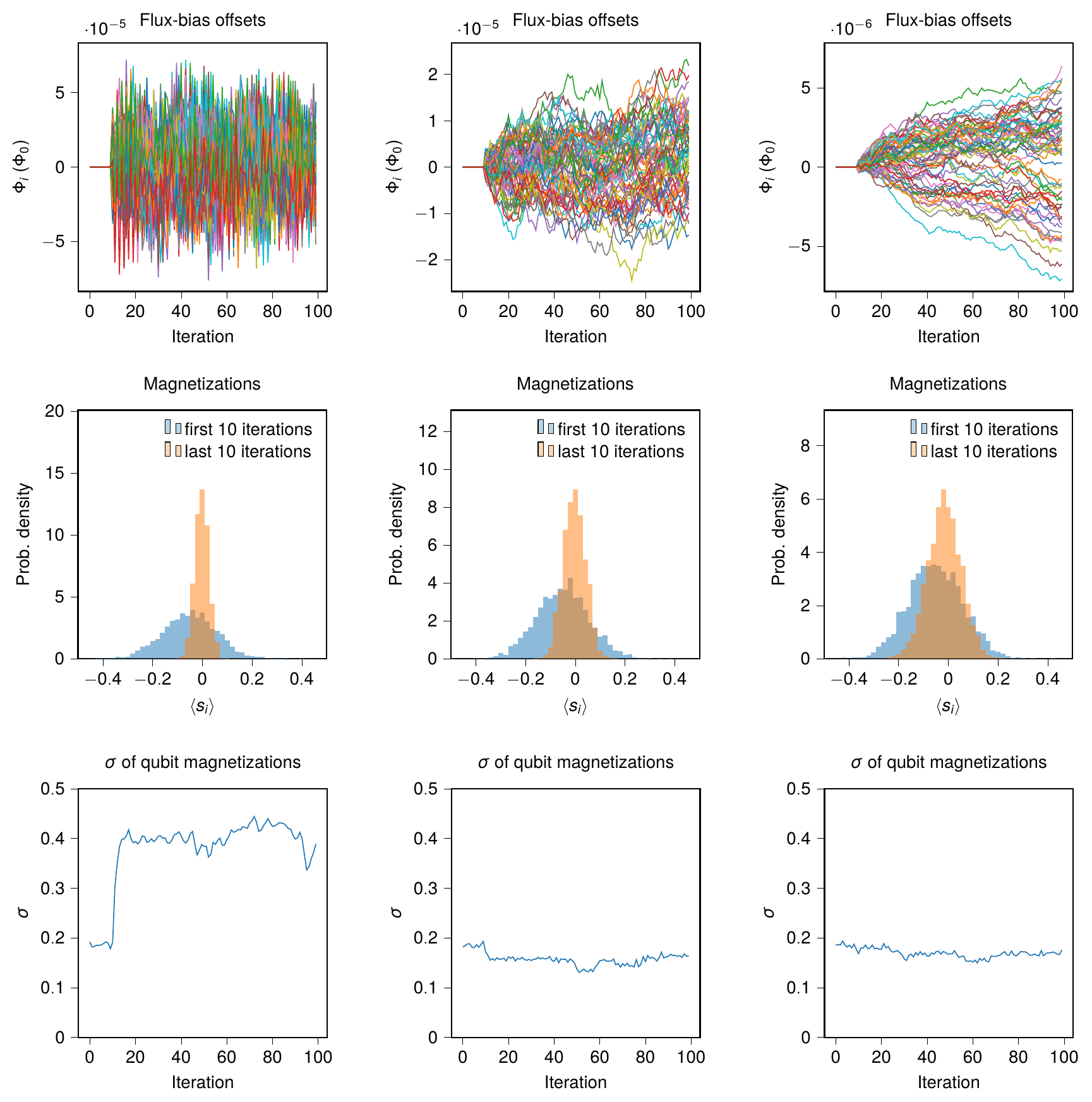}
  \caption{{\bf Balancing qubits in a FM chain with flux-bias offsets}.  Iterative correction of qubit biases is demonstrated using three step sizes $\alpha_\Phi$ for 100 iterations: $10^{-4}$ (left), $10^{-5}$ (middle), $10^{-6}$ (right).  Step size is set to zero for the first 10 iterations.  (Top) evolution of flux-bias offsets for 64 qubits in a FM chain.  (Middle) Qubit magnetization averaged over first 10 iterations and last 10 iterations.  (Bottom) Standard deviation of qubit magnetizations per iteration.
  }\label{fig:fm_fboshim}
\end{figure}

In Figure \ref{fig:fm_fboshim} we show the resulting FBOs for a single copy of the 64-qubit chain, as well as magnetization statistics.  We show experiments for three choices of $\alpha_\Phi$.  One (flux \num{1e-4}, in units of $\Phi_0$) is too large, and creates oscillations in $\Phi_i$ and $m_i$.  One (\num{1e-6}) is too small, and takes many iterations to converge.  One (\num{1e-5}) is in between, and performs well.  The choice of step size is a common concern in gradient descent applications, and we will consider automatic tuning of $\alpha_\Phi$ in a later section.  For best results, we should at a minimum ensure:
\begin{itemize}
  \item The calibration refinement appears to have converged to the vicinity of a fixed point.
  \item The parameters do not oscillate wildly.
\end{itemize}

When seeking evidence that qubit bias is improved by the FBOs, we should not just look at qubit statistics over a single QPU call, since fluctuations can be large.  Rather, we should look at the average magnetization of a qubit over multiple calls, which indicates systematic bias.  The middle row of Fig.~\ref{fig:fm_fboshim} shows the average magnetization of each qubit across the first and last ten iterations.  For each step size, the shim results in a significant improvement in variation of $m_i$ from one qubit to another.  However, the standard deviation among qubit magnetizations for {\it individual} iterations shows that the case $\alpha_\Phi=\num{1e-4}$ causes broad spreading of biases, so we need to be careful with our step sizes.

\section{Balancing spin-spin correlations}

\noindent{\bf Code reference:} \href{https://github.com/dwavesystems/shimming-tutorial/tree/main/tutorial_code}{\tt example1\_2\_fm\_loop\_correlations.py}.

Having balanced qubits at zero with linear terms with an FBO shim, we now address homogenizing the spin-spin correlations on adjacent qubits, which by symmetry should be equal for all coupled pairs.  The couplings $J_{i,i+1}$ are all nominally $-0.2$; we will fine-tune the couplings in the vicinity of this value.  This is similar to how we fine-tuned the FBOs, but with the added constraint that we do not change the average coupling.

For a given iteration we take the observed probability $f_{i,i+1}$ of the coupler being frustrated:
\begin{equation}
  f_{i,i+1} = (1+{\text{sign}}(J_{i,i+1})\langle s_is_{i_1}\rangle)/2.
\end{equation}
Let $\bar f$ denote the average frustration across all couplers in all disjoint embeddings of the chain---in general, we will compute $\bar f$ across all couplers in the union of a coupler's orbit and its opposite orbit.  We then adjust couplings based on the {\it residual frustration} $f_{i,i+1}-\bar f$:
\begin{equation}\label{eq:jshim}
  J_{i,i+1} \leftarrow J_{i,i+1}(1+\alpha_J(f_{i,i+1}-\bar f)).
\end{equation}

\begin{figure}
  \includegraphics[scale=1]{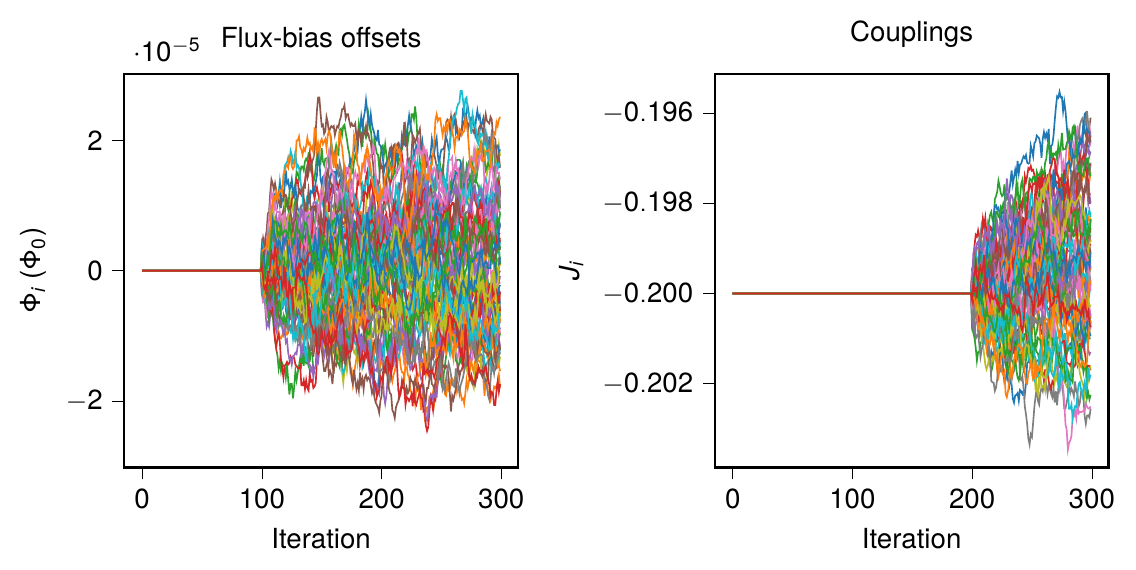}
  \caption{{\bf Balancing qubits and couplers in a FM chain with flux-bias offsets and coupler adjustments}.  This experiment is similar to that shown in Fig.~\ref{fig:fm_fboshim}, but with $\alpha_J > 0$ for the last 100 iterations.  Couplers remain distributed about the average value of $J = -0.2$.}\label{fig:fm_jshim}
\end{figure}

Fig.~\ref{fig:fm_jshim} shows data for the same experiment as Fig.~\ref{fig:fm_fboshim}, but with the ``coupler shim'' added, with $\alpha_J=0.001$.  To show the effect of the two shims, we run 100 iterations with $\alpha_\Phi=0$ and $\alpha_J=0$, then 100 iterations with $\alpha_\Phi=\num{1e-5}$ and $\alpha_J=0$, then 100 with $\alpha_\Phi=\num{1e-5}$ and $\alpha_J=0.001$.  In this particular case, the coupler shim is small but some systematic signals can be seen.  We will show more impactful cases later in the tutorial.

\newpage

\part{Worked Example: Frustrated loop}
\begin{figure}[h]{\includegraphics[scale=0.8]{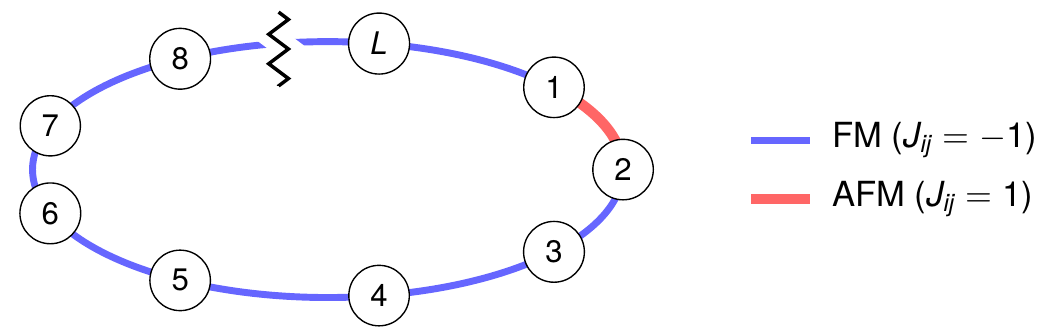}}
  \caption{Frustrated loop.}
\end{figure}

\noindent{\bf Code reference:} \href{https://github.com/dwavesystems/shimming-tutorial/tree/main/tutorial_code}{\tt example2*.py}.

Take the ferromagnetic loop considered in the previous example, and flip the sign of a single coupler $J_{1,2}$.  It is again obvious that all spins should have zero average magnetization, since there is no symmetry-breaking field (i.e., $h_i=0$ everywhere).  Less obvious is the fact that we can have two coupler orbits: one containing all FM couplers, and one containing the AFM coupler, and they are opposite.  Consequently, every coupler should be frustrated with equal probability in an ideal annealer.

\section{Finding orbits}

\noindent{\bf Code reference:} \href{https://github.com/dwavesystems/shimming-tutorial/tree/main/tutorial_code}{\tt example2\_1\_frustrated\_loop\_orbits.py}.

We can derive this fact as follows.  Flipping the sign of $s_2$, and the sign of both couplers incident to it, is a gauge transformation and, as such, will not change the probability of any coupler being frustrated in an ideal annealer.  The result of this gauge transformation is again a frustrated loop with a single AFM coupler $J_{2,3}$; note that this is equivalent to the original loop by a cyclic shift of qubit labels.  From this we can infer that $J_{1,2}$ and $J_{2,3}$ should have the same frustration probability; repeating this argument tells us that all couplers should have the same frustration probability.

\begin{figure}[h]
  \includegraphics[scale=1]{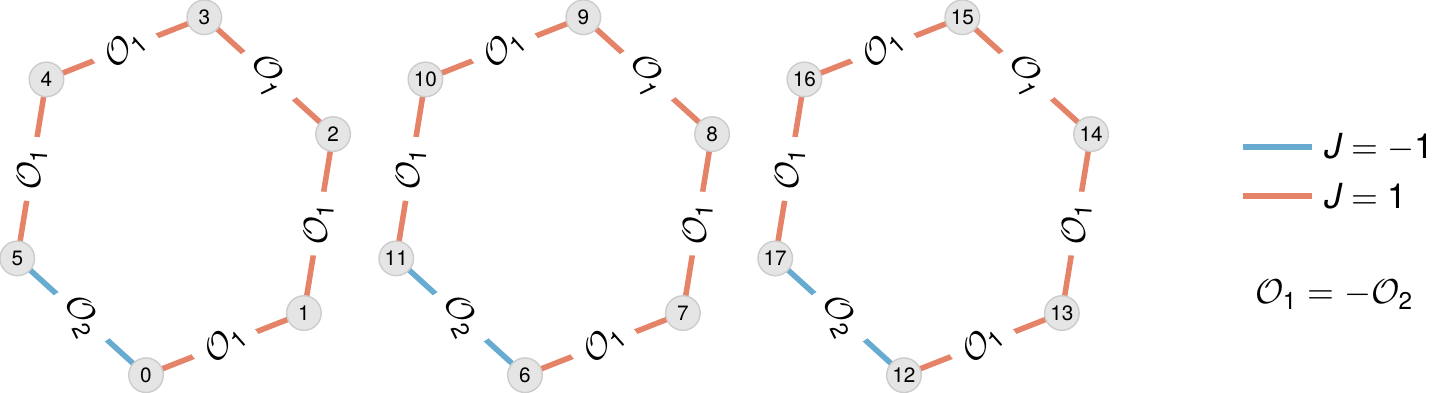}
  \caption{{\bf Coupler orbits of frustrated loops}.  The code \href{https://github.com/dwavesystems/shimming-tutorial/tree/main/tutorial_code}{\tt example2\_1\_frustrated\_loop\_orbits} constructs three disjoint frustrated loops and programmatically generates qubit and coupler orbits.  All qubits are in the same orbit.  There are two signed coupler orbits, $\mathcal O_1$ and $\mathcal O_2$, and in this example they form an opposite pair, meaning that a coupler in $\mathcal O_1$ and a coupler in $\mathcal O_2$ have opposite sign ($J=-1$ and $J=1$ in this case) but equal probability of frustration in an ideal annealer.}\label{fig:orbits_example}
\end{figure}

For more complicated examples, we would prefer to find such statistical identities programmatically as described in Section~\ref{sec:orbits}.  We do this in the file $$\href{https://github.com/dwavesystems/shimming-tutorial/tree/main/tutorial_code}{\tt example2\_1\_frustrated\_loop\_orbits.py}$$ by computing automorphisms of an auxiliary graph.  The result is a mapping $\mathcal O$ of qubits and couplers to orbits.  If spins $s_i$ and $s_j$ satisfy $\mathcal O(s_i)=\mathcal O(s_j)$, then in an ideal annealing experiment $m_i = m_j$.  Likewise, if couplers $s_is_j$ and $s_ks_\ell$ satisfy $\mathcal O(s_is_j) = \mathcal O(s_ks_\ell)$, then they have identical frustration probabilities $f_{i,j} = f_{k,\ell}$.  The code also gives us a mapping of orbits to ``opposite'' orbits, such that if spins $s_i$ and $s_j$ are in opposing orbits, $m_i=-m_j$, and if couplers $s_is_j$ and $s_ks_\ell$ are in opposing orbits then  $f_{i,j} = f_{k,\ell}$ and $J_{i,j} = -J_{k,\ell}$.

Running the code on three disjoint copies of a frustrated six-qubit loop tells us that all AFM couplers are in one coupler orbit $\mathcal O_1 = \{s_i,s_j \mid \mathcal O(s_is_j)=1\}$, and all FM couplers are in its opposite, $\mathcal O_2 = -\mathcal O_1$ (see Fig.~\ref{fig:orbits_example}).

We point out an obvious but useful fact: If we are using multiple embeddings of an Ising model, then all copies of a given qubit are in the same orbit, and all copies of a given coupler are in the same orbit.  Here we use disjoint embeddings, but they need not be disjoint: the embeddings could overlap, and be annealed in separate calls to the QPU.

\begin{figure}
  \includegraphics[scale=.9]{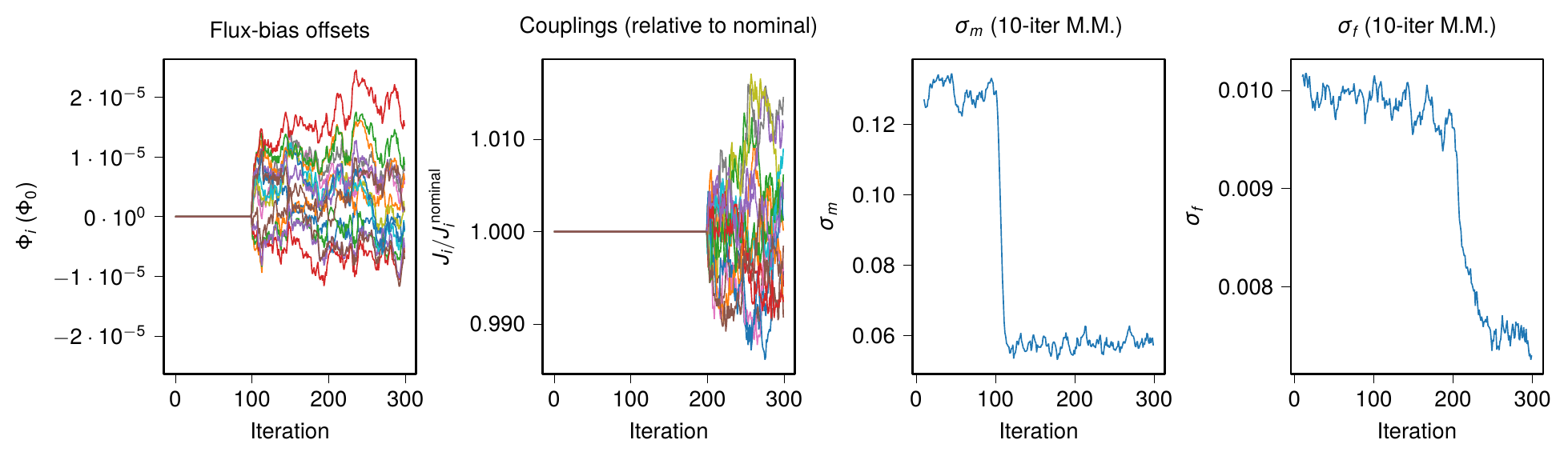}
  \caption{{\bf Shimming a frustrated loop}.  300 iterations are performed.  A flux-bias offset shim is used after iteration 100, and a coupler shim is used after iteration 200.  Nominal couplings are $\pm 0.9$.  The third panel shows the standard deviation of qubit magnetizations taken as a moving mean over 10 iterations, $\sigma_m$.  The fourth shows the corresponding quantity $\sigma_f$ for frustration probability.}\label{fig:frustloop_shim}
\end{figure}

\subsection*{Shimming}
We can now approach the frustrated loop similarly to the unfrustrated loop: all qubits should have average magnetization zero, and all couplers should be frustrated with the same probability.  Again, tuning FBOs and individual couplings helps to reduce bias in the system.  This example shows how to exploit orbits for our shim.

There is one detail worth pointing out.  In Eqs.~(\ref{eq:fboshim}) and (\ref{eq:jshim}), the terms $\bar m$ and $\bar f$ can be computed as averages over an orbit.  If we are dealing with opposing qubit orbits $\mathcal O_q$ and $-\mathcal O_q$, we can simply use $\bar m = 0$, as we do in the first example.  For opposing coupler orbits $\mathcal O_c$ and $-\mathcal O_c$, we can compute $\bar f$ across the union of the two orbits.  In this case, that means that $\bar f$ is the average frustration probability across all couplers.

Fig.~\ref{fig:frustloop_shim} shows the results of shimming FBOs and couplings for 165 parallel embeddings of a 16-qubit frustrated loop, using nominal coupling strength $|J_i|=0.9$.  Here, both components of the shim show a marked improvement of statistical homogeneity.  Taking moving means for 10 iterations at a time, we see that both $\sigma_m$ (standard deviation of qubit magnetization) and $\sigma_f$ (standard deviation of coupler frustration probability) decrease as a result of turning on the FBO shim and the coupler shim, respectively.

\subsection*{Finding orbits of an arbitrary Ising model}

\noindent{\bf Code reference:} \href{https://github.com/dwavesystems/shimming-tutorial/tree/main/tutorial_code}{\tt example2\_3\_buckyball\_orbits.py}.

Here we present an example of an Ising model that is read from a text file and run through our orbit-finding code.  The user may want to edit this code to analyze other Ising models of interest.

\begin{figure}[h]{\includegraphics[scale=0.35]{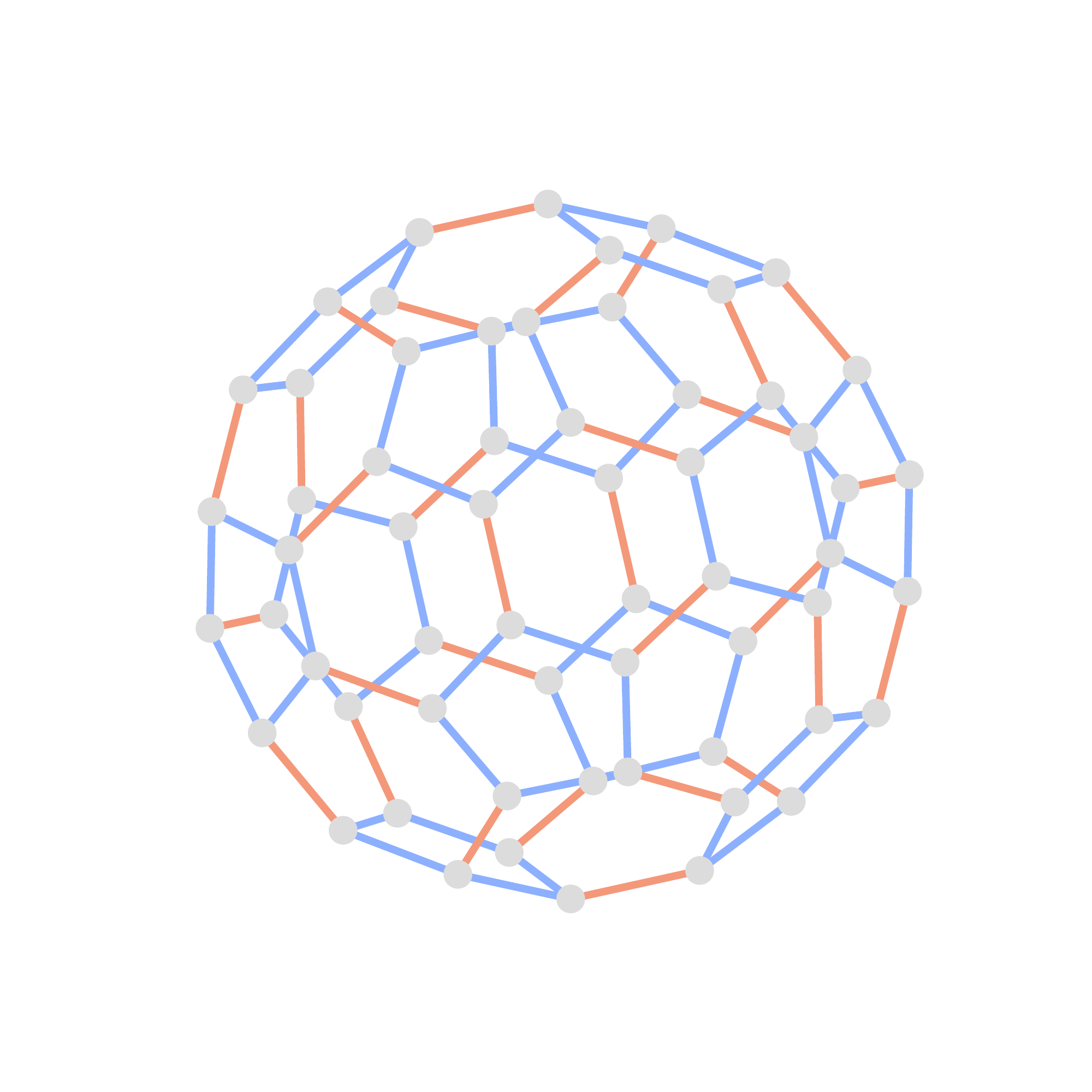}}
  \caption{An antiferromagnetic Ising model with a Buckyball graph structure. The node and edge colours encode the resulting qubit and edge orbits respectively.  All qubits are in the same orbit since the graph is vertex-transitive.  There are only two coupler orbits: those couplers sitting between two hexagons, and those sitting between a hexagon and a pentagon.}
  \label{fig:buckyball}
\end{figure}

Consider another antiferromagnetic Ising model ($J_{ij}=1$) with a \emph{Buckyball graph} as its underlying structure and no linear fields ($h_i=0$).
We apply the same methodology described in Section~\ref{sec:orbits} to find its orbits.
Fig.~\ref{fig:buckyball} visualizes the Buckyball model with its orbits labelled by text, as well as its signed Ising counterpart with coupling values encoded by colour.

\clearpage
\part{Worked Example: Triangular antiferromagnet}
\begin{figure}[h]{\includegraphics[scale=0.8]{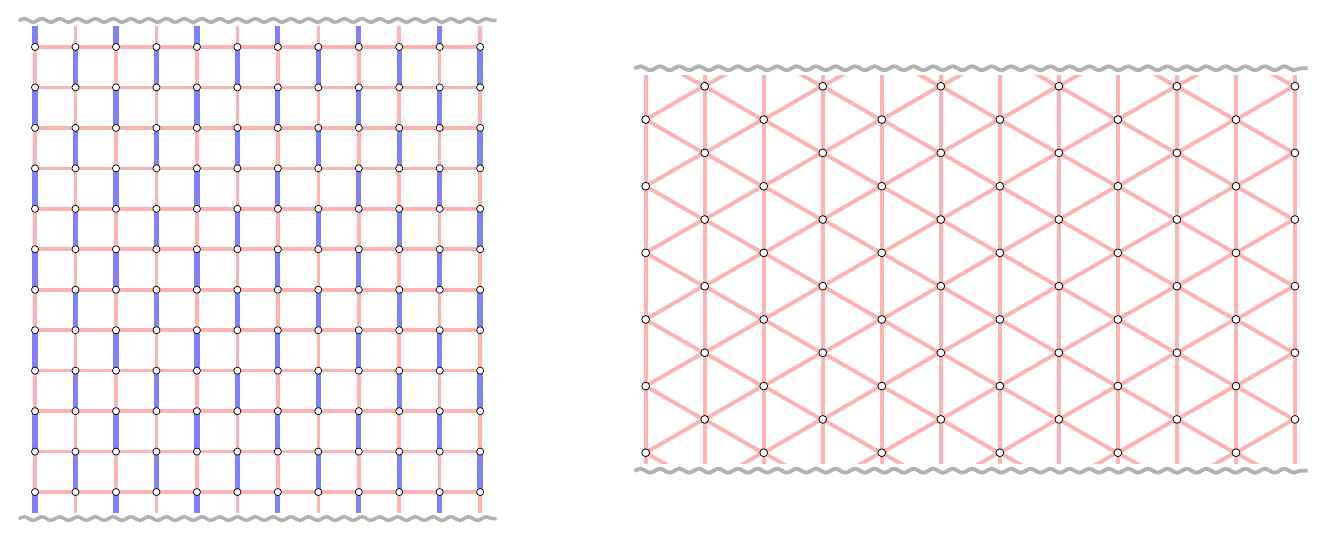}}
  \caption{A $12\times 12$ square lattice with cylindrical boundary conditions (periodic top/bottom).  Contracting two-qubit FM chains into single spins results in a triangular antiferromagnet.}
\end{figure}

\noindent{\bf Code reference:} \href{https://github.com/dwavesystems/shimming-tutorial/tree/main/tutorial_code}{\tt example3*.py}.

In the previous examples we demonstrated several key methods:
\begin{itemize}
  \item Finding qubit and coupler orbits.
  \item Homogenizing magnetizations with FBOs.
  \item Homogenizing frustration by tuning couplers.
\end{itemize}
We can now apply these tools to a nontrivial system: the triangular antiferromagnet (TAFM). This is a classic example of a frustrated 2D spin system.  Moreover, the addition of a transverse field to a TAFM leads to order-by-disorder at low temperature~\cite{Moessner2001,Isakov2003}.  For this and other reasons, including qualitative similarity to real materials, the TAFM has been simulated extensively using quantum annealers~\cite{King2018,King2022}.  We will use it as an example to showcase several concepts in calibration refinement for quantum simulation:
\begin{itemize}
  \item Truncating and renormalizing Hamiltonian terms.
  \item Simulating logical versus embedded systems.
  \item Simulating an infinite system versus faithfully simulating boundary conditions.
\end{itemize}

\section{Embedding as a square lattice}
\noindent{\bf Code reference:} \href{https://github.com/dwavesystems/shimming-tutorial/tree/main/tutorial_code}{\tt example3\_1\_tafm\_get\_orbits.py}.

In D-Wave's Advantage systems, we can minor-embed the TAFM using two-qubit FM chains.  First, we will embed a $12\times 12$ square lattice with cylindrical boundary conditions, then we ferromagnetically couple pairs of qubits with a strong coupling $J_{FM}$.  The cylindrical boundaries are very helpful in providing rotational symmetries that we can exploit in our calibration refinement methods (as in the 1D chains already studied).

The provided code uses the Glasgow subgraph solver to find embeddings of the $12\times 12$ square lattice, but note that this can take several hours.  For larger square lattices, up to $32\times 32$ or even larger depending on the location of inoperable qubits, one can inspect embeddings of smaller lattices and generalize the structure, since subgraph solvers are unlikely to be efficient at that size.  We proceed with 10 disjoint $12\times 12$ embeddings generated by the code.

In this example we will set AFM couplers to $J_{\text{AFM}}=0.9$, and all FM couplers to $J_{\text{FM}}=-2*J_{\text{AFM}}$.  Since FM couplers are very rarely frustrated in this system, we will only shim the AFM couplers.

\section{Annealing with and without shimming}

\begin{figure}
  \includegraphics[scale=.9]{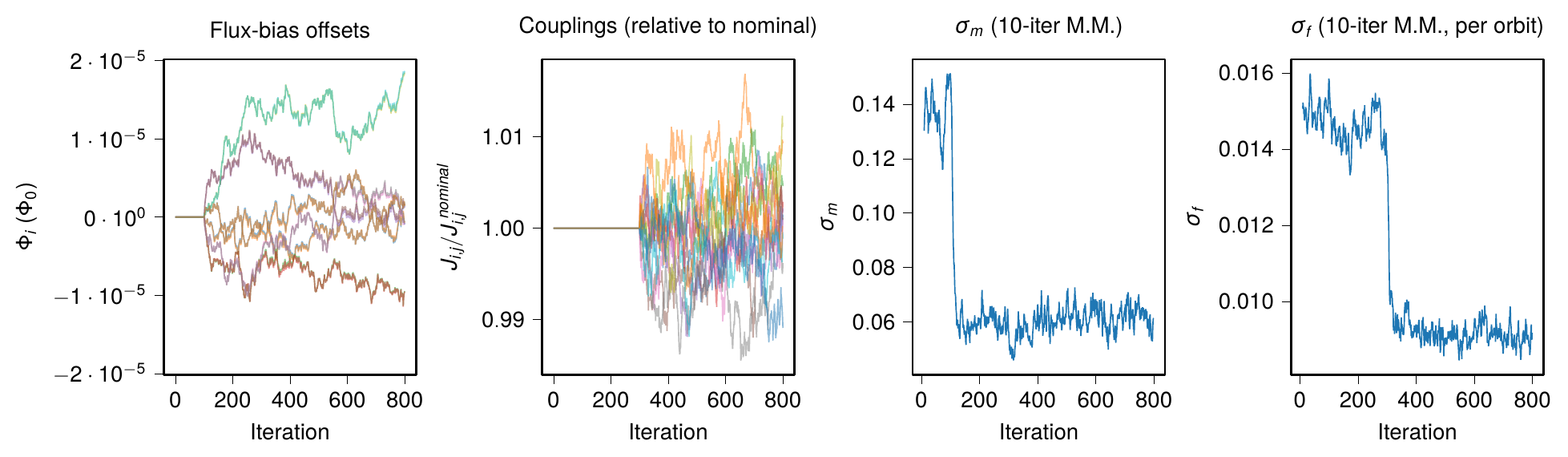}
  \caption{{\bf Shimming an embedded cylindrical triangular antiferromagnet}.  800 iterations are performed.  A flux-bias offset shim is used after iteration 100, and a coupler shim is used after iteration 300.  For clarity, we only show FBOs for 12 qubits, and couplings for 12 couplers in the same orbit.  Standard deviation of frustration probabilities, $\sigma_f$, is computed for the couplers in each orbit, and the average over all orbits is taken.}\label{fig:tafm_shim}
\end{figure}

\noindent{\bf Code reference:} \href{https://github.com/dwavesystems/shimming-tutorial/tree/main/tutorial_code}{\tt example3\_2\_tafm\_forward\_anneal.py}.

As in the previous example, we will compare performance of three methods: no shim, FBO shim only, and FBO and coupler shims together.  We perform 800 iterations, turning on the FBO shim after 100 iterations and the coupler shim after 300 iterations.
Fig.~\ref{fig:tafm_shim} shows data for this experiment, and we can see that as with the frustrated loop example, shimming improves statistical homogeneity of magnetizations and frustration.  Note, however, that there is no appreciable impact on the average magnitude of the order parameter $\langle|\psi|\rangle$.  This will change when we vary boundary conditions (see Fig.~\ref{fig:tafm_shim4}).

\section{Manipulating orbits to simulate an infinite system}

\begin{figure}
  \includegraphics[scale=.9]{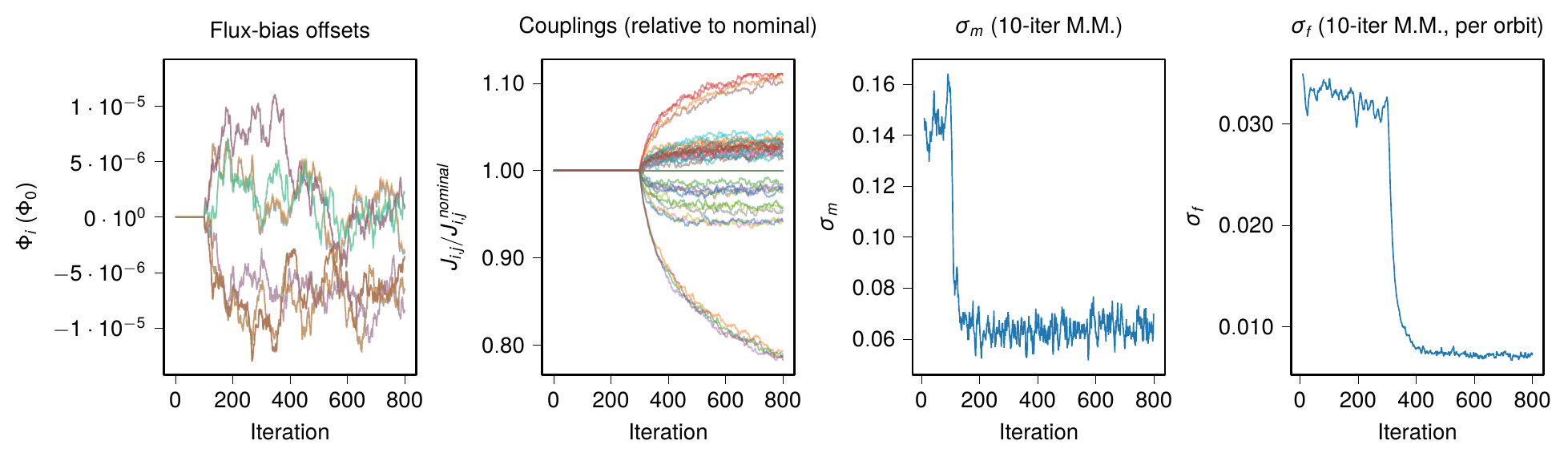}
  \caption{{\bf Shimming an isotropic, infinite triangular antiferromagnet}.  The experiment from Fig.~\ref{fig:tafm_shim} is repeated, but with all AFM couplers placed in the same orbit.  For clarity, we only show FBOs for 12 qubits, and every 5th coupling from the AFM orbit.}\label{fig:tafm_shim2}
\end{figure}

\noindent{\bf Code reference:} \href{https://github.com/dwavesystems/shimming-tutorial/tree/main/tutorial_code}{\tt example3\_2\_tafm\_forward\_anneal.py}.

The shim shown in Fig.~\ref{fig:tafm_shim} used coupler orbits for the square lattice with cylindrical boundaries, which are naturally different for couplers that are different distances from the boundary, or different orientations with respect to the boundary (and to FM chains).  But what if we want to simulate, to the extent possible, an infinite TAFM?  In that system, a coupler's probability of frustration is independent of its orientation and position, unlike in the square-lattice embedded system.  We can simulate this case by putting all AFM couplers in one orbit, and all FM couplers in a second orbit, and proceeding as before.  The coupler orbits no longer reflect the structure of the programmed Ising model, but rather the structure of the Ising model we wish to simulate.

Results for the ``infinite triangular'' shim are shown in Fig.~\ref{fig:tafm_shim2}.  This experiment is performed just like the previous one, but with the parameter
$$\text{\tt shim['type']='triangular\_infinite'}$$
instead of
$$\text{\tt shim['type']='embedded\_finite'}.$$
The coupler shim deviates significantly from nominal values (note axis scale), and has not converged even after 500 iterations.

\subsection{Truncating and renormalizing couplers}

In this code example (and others) we use an important method in the coupler shim: truncation.  Programmed couplings must be in the range $[-2,1]$, so AFM couplers must remain less than $1$, which is $1.11*J_{\text{AFM}}$.  Therefore, when couplers go out of range, we truncate them to within the range.  To avoid persistent shrinking of the couplings due to truncation, we renormalize to the correct average coupling value (0.9) before truncation---this prevents cumulative shrinkage over many iterations.

\subsection{Better initial conditions}

Looking at the data, we can see that the most reduced couplers are those on the boundary.  This suggests that if we want to simulate the infinite TAFM, we should start with a thoughtful setting of couplers.  In this case, setting the AFM couplers on the boundary to $J_{\text{AFM}}/2$ reduces the need to shim enormously.  This makes sense, since doing so maximizes the ground-state degeneracy of the classical system, as previously noted \cite{King2018}.

\begin{figure}
  \includegraphics[scale=.9]{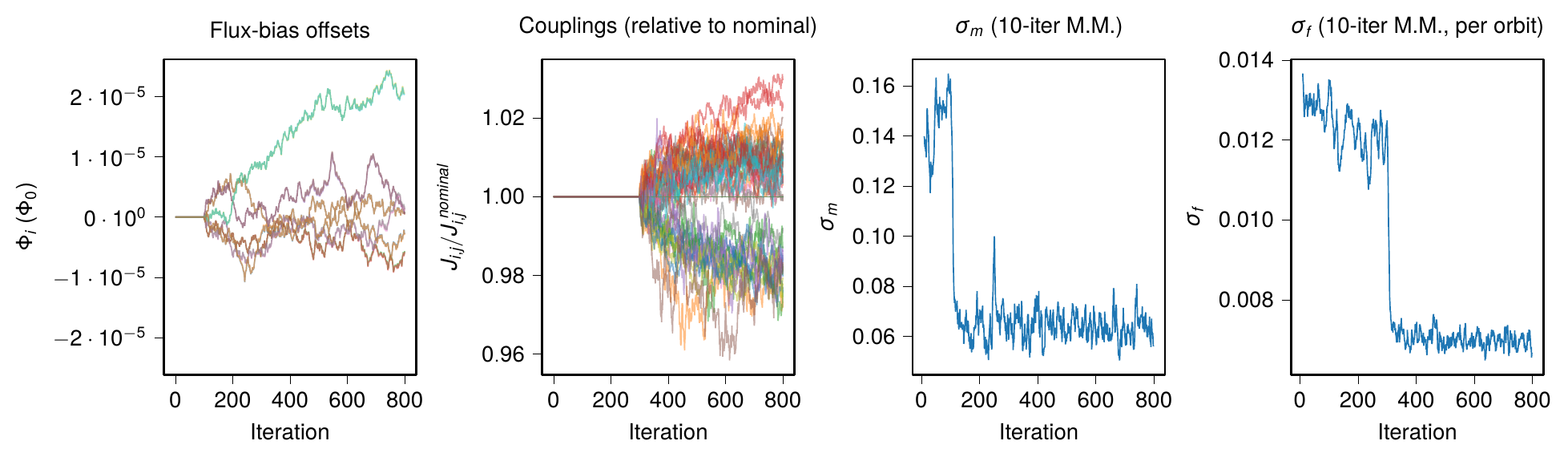}
  \caption{{\bf Shimming an isotropic, infinite triangular antiferromagnet, starting with halved boundary couplers}.  The experiment from Fig.~\ref{fig:tafm_shim2} is repeated, but with all AFM couplers on the boundary halved (to $J_{\text{AFM}}/2 = 0.45$) as an initial condition.  For clarity, we only show FBOs for 12 qubits, and every 5th coupling from the AFM orbit.}\label{fig:tafm_shim3}
\end{figure}

This shim is shown in Fig.~\ref{fig:tafm_shim3}.  The experiment is performed just like the previous one, but with the parameter
$$\text{\tt param['halve\_boundary\_couplers']=True}$$
instead of
$$\text{\tt param['halve\_boundary\_couplers']=False}.$$
We can see that now, the coupler shim only deviates a few percent from nominal, at most.

\subsection{Complex order parameter}

Order in the TAFM can be characterized by a complex order parameter $\psi$, which we define now.  Let $c:S\rightarrow \{0,1,2\}$ be a 3-coloring of the spins of the TAFM, mapping them onto three sublattices so that no two coupled spins are in the same sublattice (this coloring is unique, up to symmetries).  Then for a spin state $S$ we can define
\begin{equation}
  \psi(S) = \frac{\sqrt{3}}{N}\sum_{\ell}^N\left(s_\ell e^{c_\ell2\pi/3}  \right),
\end{equation}
where $c_i = c(s_i)$ and $i = \sqrt{-1}$.  Due to symmetries among the sublattices arising from the cylindrical boundary condition, as well as up-down symmetry of spins since $h=0$, we expect sixfold rotational symmetry (among other symmetries) in the distribution of $\psi$ in an ideal annealer.  Thus $\psi$ can serve as a good indicator of any biases in the system, as well as global ordering.

\begin{figure}
  \includegraphics[scale=.8]{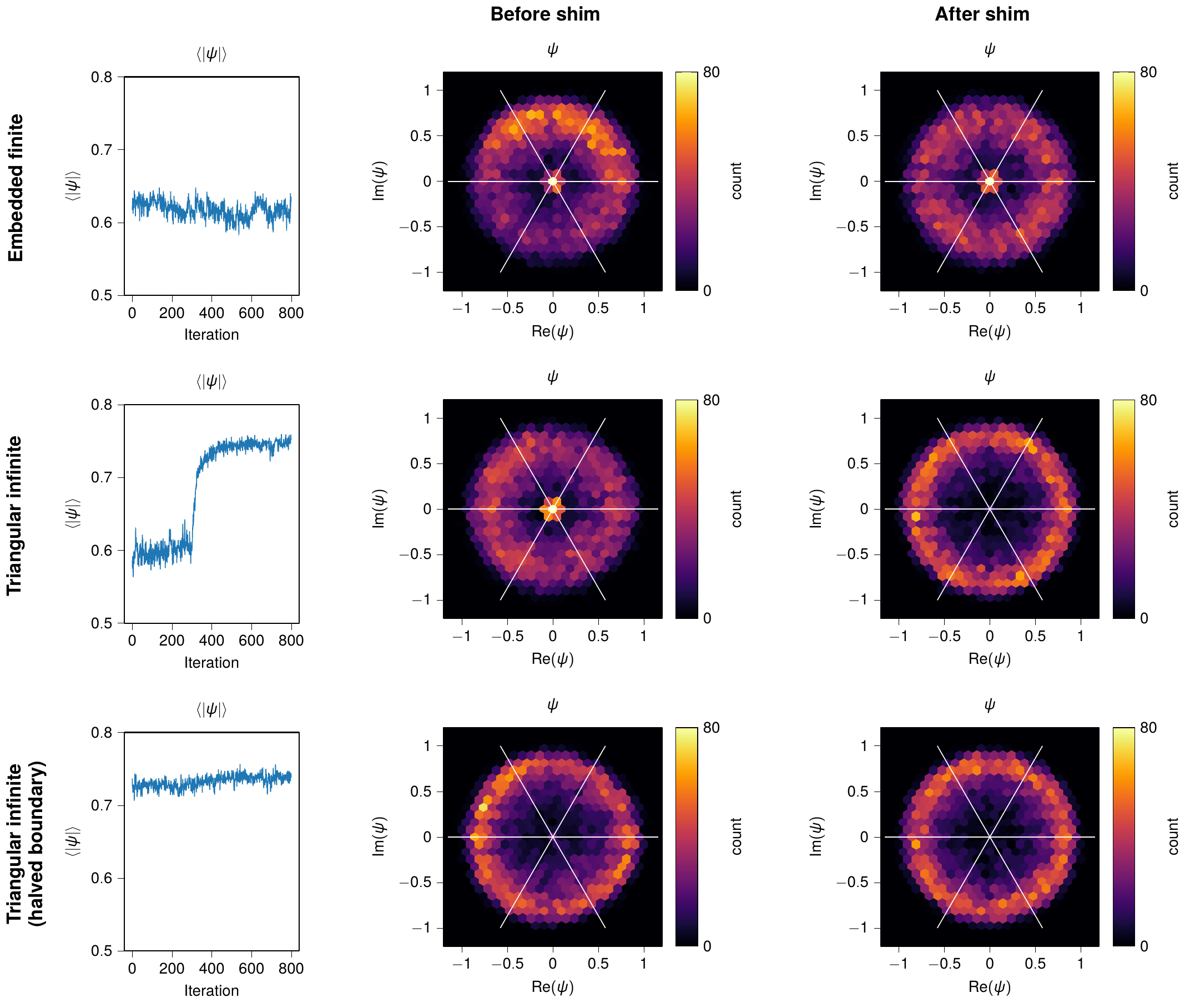}
  \caption{{\bf Complex order parameter $\psi$}.  For the three shims shown in Figs.~\ref{fig:tafm_shim}--\ref{fig:tafm_shim3}, we plot the evolution of the average magnitude $\langle \psi\rangle$, as well as complex histograms of $\psi$ (showing only data for one of the ten embeddings) before and after shimming.}\label{fig:tafm_shim4}
\end{figure}

We can use $\psi$ to compare the ``embedded finite'' shim and ``triangular infinite'' shim, as seen in Fig.~\ref{fig:tafm_shim4}.  Although we are simply forward-annealing the system, and therefore not sampling from the mid-anneal Hamiltonian, we expect the same characteristic ring histogram---without a peak near $\psi=0$---that is seen in the quantum system (cf.~\cite{King2018}~Fig.~3c).  This is seen only after the ``triangular infinite'' shim.  We mainly attribute this to the halving of the boundary couplings.  In all cases, the shim improves the theoretically expected sixfold rotational symmetry of $\psi$.

\section{Adaptive step sizes}
\noindent{\bf Code reference:} \href{https://github.com/dwavesystems/shimming-tutorial/tree/main/tutorial_code}{\tt example3\_2\_tafm\_forward\_anneal.py}.

It is often difficult or impractical to determine appropriate step sizes {\em a priori}.  Here we demonstrate a simple method for adapting step sizes based on statistics of the shim.  Note that due to noise in the QPU's surrounding environment, there is no well-defined asymptote or steady state for a shim.  However, we can loosely assume that such a state exists: we expect high-frequency fluctuations in the environment to be small compared to low-frequency fluctuations and static cross-talk.

If the step size is sufficiently small and we are sufficiently close to the steady state, we can expect fluctuations of the Hamiltonian terms (FBOs, couplers, or fields) to behave like unbiased random walks.  In an unbiased random walk with position $x(t)$ at time $t=0, 1,\ldots$, the probability distribution of $x(t)$ approaches the normal distribution with mean $0$ and variance $t$.

If the shim is far from the steady state and has a relatively small step size, the random walks will be biased in one direction, and thus the variance of fluctuations will grow superlinearly in $t$.  Finally, if the step size is very large, then it will tend to overshoot the steady state, and oscillate.  This leads to variance of fluctuations growing sublinearly in $t$.  Thus we can periodically adjust the step size of a shim as follows, using a 20-iteration lookback and a tuning term $\varepsilon=0.1$:

\begin{enumerate}
  \item For $d \leq 20$, $x(t)-x(t-d)$ is the difference between the current shim value for a term (e.g.~FBO) and the value $d$ iterations previous.  Let $X_d$ be the set of all $x(t)-x(t-d)$ for all $x$ being tuned.
  \item Find a best-fit exponent $b$ describing  $\text{var} (X_d) \propto d^b$.
  \item If $b > 1.1$, multiply the step size $\alpha$ by $1+\varepsilon$.
  \item If $b < 0.9$, divide the step size $\alpha$ by $1+\varepsilon$.
\end{enumerate}
In the example code \href{https://github.com/dwavesystems/shimming-tutorial/tree/main/tutorial_code}{\tt example3\_2\_tafm\_forward\_anneal.py}, this method is applied by setting $${\tt adaptive\_step\_size = True}.$$
This check is done every iteration, but this is not necessary.

Adaptive step sizes are so far a largely unexplored research area, and various approaches could be taken.  Using different step sizes for each orbit is certainly worth exploring; note in Fig.~\ref{fig:tafm_shim2} that different coupler orbits have hugely varying deviations from the mean.  More general frameworks like ``Adam'' \cite{Kingma2014} could also be useful in this context.

\newpage
\part{A survey of additional methods}

We have provided detailed demonstrations and free-standing Python implementations for several worked examples.  These cover the basics of calibration refinement.  Here we discuss some additional methods that have been used successfully in recent works.

\section{Shimming a system in a uniform magnetic field}

Certain Ising models in a uniform magnetic field are of interest to physicists, and these have been simulated in quantum annealers both at equilibrium \cite{Kairys2020} and out of equilibrium \cite{hysteresis}.  If we want to simulate an infinite system, we would ideally study a large system with no missing spins, and with fully periodic boundaries.  However, this is often not possible, so we wish to make the magnetization $m_i$ independent of the spin's position relative to the boundary (although it may depend on the spin's position in a unit cell of the lattice being simulated).

To deal with this, we can shim individual longitudinal field terms, $h_i$, such that all spins of a given type (i.e.~in the same position of the unit cell) are in the same orbit.  We can then shim all $h_i$ terms for each simulated field magnitude $\bar h$ that we want to study.  Then the average value of $h_i$ is forced to remain at $\bar h$ throughout the shim, perhaps with an adjustment arising from boundary spins.

To shim the case $\bar h  = 0$, we use  FBOs (as in the worked examples) instead of tuning $h_i$.

We can additionally ensure that each $h_i$ is a locally smooth function of $\bar h$ by adding a smoothing term.  For example, if $h_i$ has values $h_i^-$ and $h_i^+$ for the next lower and higher values of $\bar h$ being simulated, we can make the adjustment
\begin{equation}
  h_i \leftarrow (1-\varepsilon) h_i + \epsilon(h_i^- + h_i^+)/2,
\end{equation}
for some small constant $\epsilon > 0$.

\section{Shimming an Ising model with no symmetries}

In Ref.~\cite{QubitSpinIce}, a qubit spin ice was implemented using a checkerboard Ising model.  The system had open boundary conditions and missing spins due to inoperable qubits, so no geometric symmetries were available.  However, due to the rich automorphism group of the qubit connectivity graph (ignoring unused qubits), it was possible to generate many distinct embeddings of the same system, using different mappings of qubits to spins.  Therefore we could simulate a collection of distinct embeddings (in this case, 20) and shim in the same way we did in the worked examples.  The only difference is that in the qubit spin ice example, the embeddings are not disjoint and therefore must be sampled from using separate calls to the QPU.  However, once we have a set of samples from each embedding, we can analyze the data as though the embeddings are disjoint, whether or not this is actually the case.  The benefit remains the same: by simulating with 20 distinct embeddings, we get qubit and coupler orbits of size at least 20.

\section{Shimming a collection of random inputs}

In Ref.~\cite{King2022b}, shimming was used to study {\it spin-glass ensembles}---collections of random problems with certain parameters.  As we have seen, we can spend hundreds of iterations shimming a single problem, and this becomes impractical when studying ensembles of thousands of instances.

The approach used was to exploit a common symmetry: all problems in the ensembles had $h=0$.  Shimming the couplers was abandoned as being impractical for such a large set of inputs.  Shimming FBOs, however, is straightforward.  By cycling through 300 spin-glass realizations using the same set of qubits and couplers, simulating each realization several times, it is possible to combine the work and arrive at a good set of FBOs that mitigates the majority of systematic offsets.

\section{Shimming anneal offsets for fast anneals}

As described in the Supplementary Materials to Ref.~\cite{King2022}, D-Wave quantum annealing processors have recently demonstrated the capacity to anneal much faster than currently generally available, at an anneal time of 10 nanoseconds or less \cite{King2022, King2022b}.  This speed exceeds the ability of the control electronics to synchronize the annealing lines (eight in the Advantage processor, four in D-Wave 2000Q\texttrademark ) satisfactorily.  Therefore, frustration statistics can be used to infer which lines are out of sync with the others, and in which direction.  Anneal offsets, which allow individual qubits to be annealed slightly ahead of or behind other qubits, were used to synchronize the qubits on each annealing line.  These fast anneals are not currently generally available, but they may be in the future.

\newpage
\part{Additional tips}

\section{Making calibration refinement more efficient}

As we have seen, shimming can take many iterations to converge.  Naively repeating the process across many combinations of parameters (e.g., annealing time, energy scale, etc.) can be extremely time consuming.  However, there are ways to improve the efficiency of the process.  Here we outline some important things to bear in mind.

\subsection{Adjustments are often continuous functions of other parameters}

If we determine a set of adjustments for a given experiment, then slightly vary some parameters of the experiment, we can generally expect that the adjustments will not change much.  For example, FBOs and coupling adjustments are expected to vary smoothly as functions of annealing time, energy scale, and various perturbations to the system (for example the ratio between FM and AFM couplers in an embedded triangular antiferromagnet).  An important example is the annealing parameter $s$, in cases where we simulate a system at $0 \ll s \ll 1$ (\cite{King2018, King2021, King2022} etc).

As an example of how this can help speed up a shim, if we double the annealing time, FBOs and coupling adjustments will remain relatively stable.  Thus, rather than starting our shim anew from the nominal Hamiltonian, we can start from an adjusted Hamiltonian that was determined using similar parameters.  One could go further than this, and extrapolate or interpolate based on multiple values.

\subsection{Predictable adjustments should be programmed into the initial Hamiltonian}

As shown in Figs.~\ref{fig:tafm_shim2} and \ref{fig:tafm_shim3}, starting with halved boundary couplings can immediately bring the couplings close to their converged values.  If we are aware of such adjustments, using them as initial conditions can make shims converge far faster.

\section{Damping shim terms}

It is sometimes useful to gently encourage a shim to remain close to the nominal values, for example to prevent drifting Hamiltonian terms.  This issue can be particularly important near a phase transition, where statistical fluctuations can be very large.  Drift can be suppressed by adding a damping term to the shim.  For example, we can set a damping constant $0 \leq \rho \leq 1$, and after every iteration we can move each coupler $J_{ij}$ towards its nominal value $\hat J_{ij}$:
\begin{equation}
  J_{ij} \rightarrow J_{ij} - \rho(J_{ij}-\hat J_{ij}).
\end{equation}
Doing this can discourage random fluctuations, but can also lead to undercompensation of biases.  It is only recommend to use damping when the shim is otherwise badly behaved.

\newpage
\part{Conclusions}

In this document we have presented several basic examples that introduce the value of calibration refinement or ``shimming'' in quantum annealing processors.  These methods should be applied to any detailed study of quantum systems in a quantum annealer, and will generally provide a significant improvement to the results.  Depending on the sensitivity of the system under study, these methods can mean the difference between an unsuccessful experiment and an extremely accurate simulation.

We have provided fully coded examples in Python, which should be easy to generalize and adapt.  As part of these examples, we include methods for embedding many copies of a small Ising model in a large quantum annealing processor.  This is a valuable and straightforward practice that can enormously improve both the quantity and the quality of results drawn from a single QPU programming.

Another important perspective, which has been introduced here for the first time, is the notion of constructing an auxiliary Ising model and using automorphisms of it to infer qubit and coupler orbits automatically.  We encourage users to experiment with this method and report on any challenges or benefits found.

The examples in this document are written for use in an Advantage processor, but are not specific to that model, or even to D-Wave quantum annealers in general.  These results may prove useful in diverse analog Ising machines, both quantum and classical.

\section*{Acknowledgments}

The authors are grateful to Ciaran McCreesh for help with the Glasgow Subgraph Solver, and to Hanjing Xu, Alejandro Lopez-Bezanilla, and Joel Pasvolsky for comments on the manuscript.

\bibliography{paper}

\end{document}